# THE OPEN PROBLEM OF FINDING A GENERAL CLASSIFICATION OF GEODETIC GRAPHS

Carlos E. Frasser

This note describes some open problems that can be examined with the purpose of gaining additional insight of how to solve the problem of finding a general classification of geodetic graphs.

## 1. BACKGROUND

A geodetic graph is one in which every pair of its vertices has a unique path of shortest length. Examples of geodetic graphs are trees, complete graphs and odd-length cycles. In the 1962 edition of his book on graph theory, mathematician Oystein Ore posed the characterization problem of geodetic graphs. Over the years, despite good attempts by several authors to accomplish the task, a characterization has proven to be extremely difficult. There have been several approaches to tackle the problem. Initially, the properties of known geodetic graphs such as complete graphs were studied. This allowed for a full characterization of geodetic graphs homeomorphic to a complete graph on four vertices and also a complete description of planar geodetic graphs. Then, the focus of attention switched to the class of complete graphs with more than four vertices. After several attempts, geodetic graphs homeomorphic to a complete graph on $n$ vertices were fully characterized and, at the same time, the first descriptions of construction, mostly involving section subgraphs of complete graphs, started to be an inherent part of the problem of characterization. The fact that the known Moore graphs of diameter 2, that is, the cycle of length five, the Petersen graph and the Hoffman-Singleton graph, turned out to be geodetic, a characterization of the geodetic graphs of diameter 2 with no cliques was also obtained. As a result, geodetic graphs of diameter 2 with and without cliques were examined and a full characterization included graphs constructed from a generalization of affine planes and Moore graphs of diameter 2. Next, the attention was focused on geodetic graphs of diameter 3 and several aspects of their structure were studied. After these developments, several constructions of geodetic graphs of diameter 4 and 5 were described and examined in detail using balanced incomplete block designs (BIBDs.) Despite these efforts, a full characterization is still elusive. We think that a complete structural description of geodetic graphs is the direct path to their characterization. The problem of the description of how geodetic graphs are structurally constructed was researched by this author in a couple of past papers. Any geodetic graph has a similar structural description to that of a Moore graph. This permitted us to create a new approach to the problem of generating the class of all geodetic graphs homeomorphic to a given geodetic one. With the help of combinatorics and an algorithmic procedure, we were able to construct new classes of geodetic graphs for every value of the diameter that we were considering. But then we realized that the number of graphs generated after each iteration of the devised algorithm followed a clear pattern. This suggested the idea that beside generating the graphs we were to construct, our algorithm allowed us to find their general number. Therefore, beside the problem of construction, now the problem of enumeration is believed to play an important role in the whole picture of the problem of characterization. Recent developments in the research of geodetic graphs using tools of theoretical computer science emphasize the problem of their structural description, which, we believe, could be the key approach for a full characterization. Finally, this note is the result of our efforts of reaching a full characterization and outlines the unsolved problems that can be examined with the purpose of gaining additional insight of how to solve the open problem of finding a general classification of geodetic graphs.

## 2. INTRODUCTION

As mentioned before, Ore [9] posed the question of characterizing geodetic graphs, which is still unresolved. Later on, beside the problem of characterization [11, 12, 14, 15], several other research problems related to geodetic graphs were added, namely: the problem of their construction [6, 10, 12, 13], the problem of their enumeration [6, 16, 17] and the interesting problem of their application not only to the topological design of optimal computer systems and networks [4, 5], but to other branches of mathematical sciences [7]. Characterization, enumeration and the general classification of geodetic graphs are open problems that have become a very important research topic due to the fact that geodetic graphs can be applied to the solution of other relevant problems in computer science [1, 2, 3]. Because of the increasing difficulty finding new approaches for their characterization and new techniques for their construction, a simpler approach chosen by some authors has been to try to deal with them by specific classes, which has contributed to establish some generalizations related to the description of their characterization [6, 15] and has allowed to enumerate some of those classes [6, 16, 17]. All of those previous research efforts have contributed to the idea of trying a general classification by constructing and enumerating them algorithmically [2, 6, 8]. It is believed that these new relevant ideas could give important insight of how to classify them once and for all. Hopefully, it can be done soon.

The following are open problems related to the topics mentioned above. It is thought that the majority of them are interrelated and their solution will allow to understand better the whole picture of the general classification problem of geodetic graphs. For that purpose, we need some definitions.

The set of vertices and edges of a graph $G$ are denoted $V(G)$ and $E(G)$, respectively. A *path* from $v_0$ to $v_n$ is a sequence $v_0v_1\ldots v_n$ of different vertices of $G$ and is denoted $P(v_0, v_n)$. The number of edges of a path $P$ determines the length of this path and is represented by $|P|$. A cycle is a sequence $v_0v_1\ldots v_nv_0$ of different vertices of $G$.

The length of a shortest path connecting vertices $u$ and $v$ in $G$ represents the *distance* between these two vertices. The greatest distance between any pair of vertices of $G$ is called the *diameter* of $G$, which is denoted $d(G)$. A *geodesic* in $G$ is a path of shortest length whose endpoints are two different vertices $u$ and $v$ of $G$. A graph $G$ is *geodetic* if it contains a unique geodesic between any pair of its different vertices $u$ and $v$.

The number of edges incident to $v$ is called the *degree* of vertex $v$ and is denoted $deg(v)$. $G$ is said to be regular of degree $k$ if every vertex of $G$ has equal degree $k$. A vertex for which $deg(v) \geq 3$ is called a *node*. A path, whose only nodes are its endpoints $u$ and $v$, is called a *segment*. A graph $G$ is *strongly regular* if it is $k$-regular and there exist integers $\lambda$ and $\mu$ such that for all pairs of different vertices $u$, $v$ of $G$, the following satisfies: i) if $u$ and $v$ are adjacent, there are exactly $\lambda$ vertices adjacent to both $u$ and $v$ and ii) if $u$ and $v$ are nonadjacent, there are exactly $\mu$ vertices adjacent to both $u$ and $v$.

Graph $G_1$ is *homeomorphic* to graph $G_2$ iff both $G_1$ and $G_2$ can be obtained by insertion of vertices onto their edges. Note that each graph is homeomorphic to itself. A graph in which every pair of its vertices is adjacent is called *complete*. The complete graph on $n$ vertices is denoted $K_n$. A *clique* is defined as a maximal complete subgraph $K_n$, $n \geq 3$, that is contained in no larger complete subgraph.

The minimum number of vertices whose deletion (implies also the deletion of the edges incident to the deleted vertices) disconnects $G$ is called *vertex connectivity* of a graph $G$. A *block* is a graph whose vertex connectivity is greater than 1.

A regular graph $G$ of degree $k$ and diameter $d$ is called a *Moore graph* if

$$|V(G)| = 1 + k \sum_{j=1}^{d}(k-1)^{j-1} \qquad (1)$$

Any geodetic block of diameter 2 that does not contain cliques $K_n$, $n \geq 3$, is a Moore graph of the following types: a cycle of length 5 ($k = 2$), the Petersen graph ($k = 3$), the Hoffman-Singleton graph ($k = 7$). Case $k = 57$ is a block whose construction is still open [14].

## 3. A DESCRIPTION OF THE OPEN PROBLEMS

### A. Open Problems Related to Characterization

It is known that the following three conditions are sufficient for a homeomorphic graph $G$ to the Petersen graph to be geodetic:

(1) Each path composed of two successive segments, one followed by the other, is a geodesic in $G$.
(2) Each cycle of $G$ containing five segments has odd length.
(3) All cycles of $G$ containing six segments have equal length.

More generally, the following three conditions are sufficient for a homeomorphic graph $G$ to any Moore graph of diameter $d$ ($k \geq 3$) distinct from $K_n$, $n \geq 5$, to be geodetic:

(1) Each path of $G$ composed of $d$ successive segments, one followed by the other, is a geodesic in $G$.
(2) Each cycle of $G$ containing $2d + 1$ segments has odd length.
(3) All cycles of $G$ containing $2d + 2$ segments have equal length.

It is conjectured that the three conditions of each of the two previous statements are not only sufficient, but also necessary [6].

### B. Open Problems Related to Construction and Enumeration

Assuming that the previous conjectures of characterization are true, a new technique to generate the class of all geodetic graphs homeomorphic to a given geodetic graph that deals with the construction of the so-called geodetic system of Diophantine equations associated with the given geodetic block, each vertex of it having degree $k \geq 3$, has allowed to construct and enumerate them algorithmically [6]. As a result, several open problems arise

1. Can be proved rigorously that the general number of all geodetic graphs of diameter $d \geq 2$ homeomorphic to the Petersen graph is $_{(d+3)}C_5$ (the number of combinations of $d + 3$ elements taken 5 at the time)?
2. Is it possible to devise a computer program to be run in a modern computer in real time that allows to enumerate the set of all natural solutions of the geodetic system of Diophantine equations associated with the Hoffman-Singleton graph and to generate and find the general number of all geodetic graphs of diameter $d \geq 2$ homeomorphic to it as was done for the Petersen graph in [6]?
3. Does there exist any sort of hidden structure contained within the distribution of the variables of the geodetic system of Diophantine equations associated with $K_4$, the Petersen graph, and the Hoffman-Singleton graph that is common to the three systems and allows to construct the system to generate the class of all geodetic graphs homeomorphic to the undiscovered Moore graph ($k = 57$) including itself?
4. Do there exist infinitely many strongly regular geodetic graphs, or any strongly regular geodetic graphs that are not Moore graphs?

### C. Open Problems Related to Computer Networks

Discrete optimization, which is defined as a mathematical discipline that comprises not only combinatorial optimization, but also general (mixed-)integer optimization problems with no special combinatorial structure, and graph enumeration are theoretical topics very close related to the applied problem of creating complex computer networks of optimum performance, that is, computer networks with optimum indices of their operation quality such as systemic hierarchy, diameter, vertex connectivity, reliability, cost, and flexibility of topological change, among others [4, 5]. It is known that certain classes of graphs (geodetic graphs) are better suited than others when dealing with the problem of obtaining optimum indices of the quality of network operation. That quality of work can be significantly improved by changing the topology of connections, which in the case of homeomorphism can be done by knowing the exact number of graphs of given diameter $d$ that represents the same network topology [5]. An open problem is to research how the change of topology of connections influences the quality of network performance. Another open problem consists in determining a single model that enables to consider many different characteristics of the quality of network operation and an adequate technique to optimize them.

### D. Open Problems Related to Computer Science

It is known that the general classification problem of geodetic graphs is related to the problem of the study of rewriting systems, which, at the same time, connects abstract algebra and theoretical computer science [2, 3]. As a result, a new contribution to a better understanding of the description of geodetic graphs consists in proving several conjectures, among them, the following one:

"Let $G$ be a locally-finite simple geodetic graph, and $t \in \mathbb{N}$. If $C$ is an embedded cycle of diameter exceeding $t$ that has minimal length among all such paths in $G$, then $C$ contains a geodesic subpath of length $t + 1$."

Finally, a contribution to a better understanding of the structure of geodetic graphs applies mathematical methods of graph drawing to the problem of finding the number of crossings of minimal length paths (geodesics) in a geodetic graph. A couple of open problems are posed regarding this issue [1].